\documentclass[aps,prl,print,twocolumn,groupedaddress,showkeys]{revtex4-2}
\usepackage{graphicx}
\usepackage{amssymb}
\usepackage{amsfonts}
\usepackage{amsmath}
\usepackage{natbib}
\usepackage{hyperref}
\usepackage{color}
\usepackage{bm}
\usepackage{mathrsfs}
\usepackage{amsthm}

\begin{document}

\title{Realization of a non-Hermitian Haldane model in circuits}
\author{Rujiang Li$^{1}$}
\thanks{Corresponding author: {rujiangli@xidian.edu.cn}}
\author{Wencai Wang$^{1}$}
\author{Xiangyu Kong$^{1}$}
\author{Bo Lv$^{2}$}
\author{Yongtao Jia$^{1}$}
\author{Huibin Tao$^{3}$}
\thanks{Corresponding author: {coldfire2000@mail.xjtu.edu.cn}}
\author{Pengfei Li$^{4,5}$}
%\thanks{Corresponding author: {lipf@tynu.edu.cn}}
\author{Ying Liu$^{1}$}
%\thanks{Corresponding author: {liuying@mail.xidian.edu.cn}}

\begin{abstract}

The Haldane model is the simplest yet most powerful topological lattice model exhibiting various phases, including the Dirac semimetal phase and the anomalous quantum Hall phase (also known as the Chern insulator). Although considered unlikely to be physically directly realizable in condensed matter systems, it has been experimentally demonstrated in other physical settings such as cold atoms, where Hermiticity is usually preserved. Extending this model to the non-Hermitian regime with energy non-conservation can significantly enrich topological phases that lack Hermitian counterparts; however, such exploration remains experimentally challenging due to the lack of suitable physical platforms. Here, based on electric circuits, we report the experimental realization of a genuine non-Hermitian Haldane model with asymmetric next-nearest-neighbor hopping. We observe two previously uncovered phases: a non-Hermitian Chern insulator and a non-Hermitian semimetal phase, both exhibiting boundary-dependent amplifying or dissipative chiral edge states. Our work paves the way for exploring non-Hermiticity-induced unconventional topological phases in the Haldane model.

\end{abstract}

\affiliation{$^1$National Key Laboratory of Radar Detection and Sensing, School of Electronic 
Engineering, Xidian University, Xi'an 710071, China}

\affiliation{$^2$Key Laboratory of In-Fiber Integrated Optics of Ministry of Education, 
College of Physics and Optoelectronic Engineering, Harbin Engineering University, Harbin 150001, Heilongjiang Province, China}

\affiliation{$^3$School of Software Engineering, Xi'an Jiaotong University,
Xi'an, China}

\affiliation{$^4$Department of Physics, Taiyuan Normal University, Jinzhong, 030619,
China}

\affiliation{$^5$Institute of Computational and Applied Physics, Taiyuan Normal University,
Jinzhong, 030619, Shanxi, China}


\maketitle

\noindent \textbf{Introduction}

\noindent  The Haldane model describes a two-dimensional (2D) graphene-like lattice with band structure properties determined by nearest-neighbor (NN) and next-nearest-neighbor (NNN) couplings \cite{PRL61-2015}. Historically, the Haldane model laid the foundation for the discovery of topological phases of matter, including anomalous quantum Hall insulators (also known as Chern insulators) and quantum spin Hall insulators (also referred to as topological insulators) \cite{RMP82-3045, RMP83-1057}. Although initially proposed for electrons and considered a purely quantum model, the Haldane model has since been applied to classical wave systems, giving rise to various emerging fields such as topological photonics, topological acoustics, and topological circuits \cite{nphoton8-821, RMP91-015006, nphoton11-763, PRL114-114301, NRP1-281, nphys14-875, apr2-2100013, LSA9-130, PR1093-11, nanophotonics10-425, PQE55-52}.

In the phase diagram of the Haldane model, there are typically three distinct phases: trivial insulators, 2D Dirac semimetals with either paired or unpaired Dirac points, and Chern insulators \cite{PRL61-2015}. From the perspective of symmetry analysis, the phase diagram is primarily influenced by inversion symmetry ($P$), which is generally determined by the energy offset between the two sublattices, and time-reversal symmetry ($T$), which is typically dictated by the strength of the local magnetic flux. By breaking $P$ while preserving $T$, the Haldane model can give rise to trivial insulators, which are later recognized as valley-Hall insulators \cite{apr2-2100013}. Conversely, breaking $T$ while preserving $P$ can lead to the emergence of Chern insulators. Between these two gapped phases lie the gapless Dirac semimetal phases.

A third crucial symmetry consideration in the Haldane model is Hermiticity, which refers to the property that a Hamiltonian equals its conjugate transpose. Conventionally, the Haldane model is considered Hermitian, as most condensed matter systems conserve energy. However, recent developments in non-Hermitian physics have drawn researchers' attention to the search for topological phases within the non-Hermitian Haldane model, which exhibit unique characteristics not found in Hermitian systems \cite{science363-eaar7709, RMP93-015005, nphys17-9, nanophotonics9-547}. 
In the Haldane model, there are typically two approaches to introduce non-Hermiticity. The first approach involves adding on-site gains or losses, which can interact with parity-time ($PT$) symmetry and exceptional points \cite{PRB98-165129, PRR2-013387}. It is important to note that introducing a uniform lossy background to the Haldane model does not lead to novel non-Hermitian topological phases, rendering this approach trivial. The second approach is to generalize the antisymmetric next-nearest-neighbor (NNN) hopping into an asymmetric one \cite{PRB100-081401}. In this configuration, the chiral edge states along two parallel edges can possess a pair of conjugated energies, implying the existence of one amplifying and one dissipative topological edge state. However, to date, neither of these nontrivial non-Hermitian Haldane models has been experimentally reported.

\begin{figure*}[tbp]
\includegraphics[width=14.8cm]{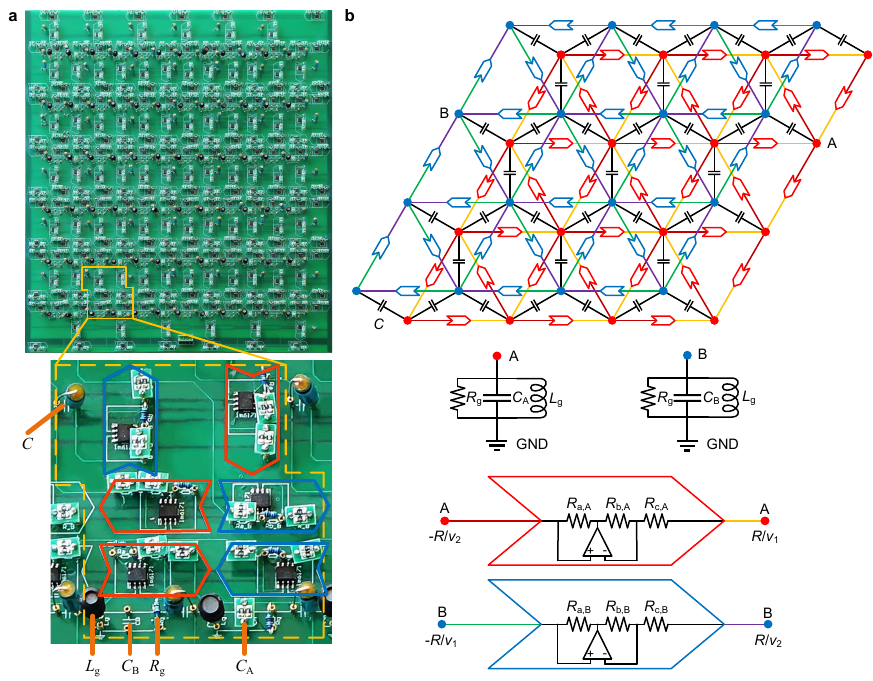}
\caption{\textbf{Circuit realization of a non-Hermitian Haldane model.} 
\textbf{a}, The PCB of the fabricated circuit. The honeycomb lattice in (\textbf{b}) is transformed into a brick wall structure, with an enlarged view showing the circuit board cutout of one unit cell. 
\textbf{b}, Schematic of the circuit. Nodes A (red) and B (blue) are connected to grounded $LC$ resonators with parameters $L_{\text{g}} = 68~\mu \text{H}$, $C_{\text{A}} = 20~\text{nF}$, and $C_{\text{B}} = 0$. The nearest-neighbor (NN) nodes are interconnected by coupling capacitors with $C = 10~\text{nF}$. The next-nearest-neighbor (NNN) nodes are connected through negative impedance converters (NICs), which provide $R_{\text{A},\text{B}} = -R/\nu_{2,1}$ for the left side and $R_{\text{A},\text{B}} = R/\nu_{1,2}$ for the right side, with $R = 1~\text{k}\Omega$.
}
\label{fig1}
\end{figure*}

The grand challenge of implementing the non-Hermitian Haldane model lies in simultaneously realizing nonreciprocal next-nearest-neighbor (NNN) couplings and controlling non-Hermiticity. Previous systems have typically been able to achieve only one of these conditions. Conversely, electric circuits have recently emerged as a promising platform for implementing various topological phenomena, particularly their non-Hermitian counterparts, due to their precise control over couplings (both reciprocal and nonreciprocal) and on-site potentials using passive or active lumped components. Notable examples include higher-order topological states, nonlinear topological edge states, non-Hermitian skin effects, and many others \cite{PRB99-020304, PRB100-201406, LSA9-145, nphys14-925, PRB102-104109, PRB102-100102, PRR2-022028, nelectron1-178, ncommun10-1102, PNAS118-e2106411118, ncommun11-2356, nphys16-747, ncommun12-7201, FOP20-14203, FOP18-33311}. However, realizing a non-Hermitian Haldane circuit model has thus far remained elusive.

\begin{figure*}[tbp]
\includegraphics[width=14cm]{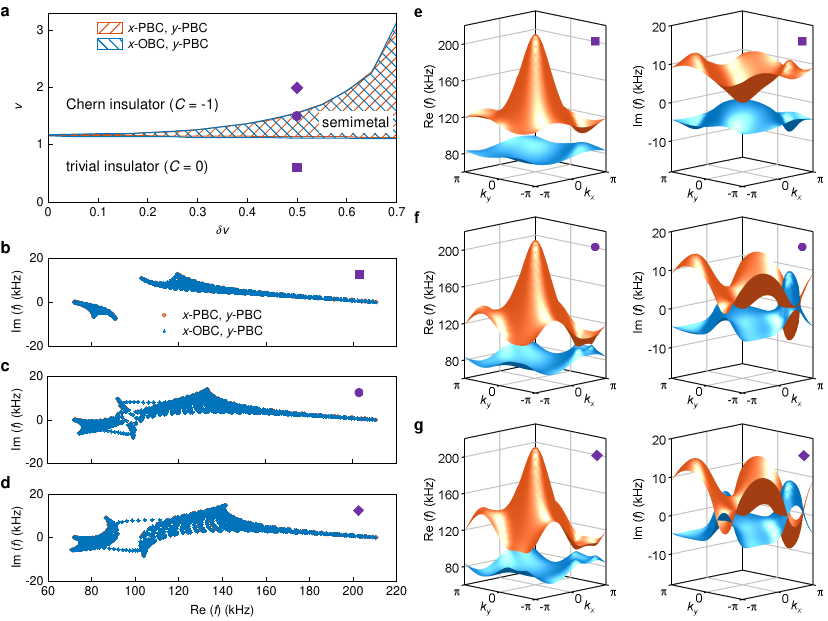}
\caption{\textbf{Phase diagram of the circuit lattice.} 
\textbf{a}, Phase diagram of the non-Hermitian Haldane model, where the Chern insulator and trivial insulator regimes are characterized by Chern numbers $C = -1$ and $C = 0$, respectively. The gapless regime (shaded area) for the semimetal corresponds to a gapless line gap defined by $\text{Re} \left( f_{0} \right) = 0$. 
\textbf{b}-\textbf{d}, Complex frequency spectra for (\textbf{b}) the trivial phase with $\nu = 0.6$, (\textbf{c}) the gapless phase with $\nu = 1.5$, and (\textbf{d}) the nontrivial phase with $\nu = 2$. In panels (\textbf{a})-(\textbf{d}), the red and blue colors represent results calculated from the circuit lattices with PBC and OBC in the $x$-direction, respectively, while PBCs are applied in the $y$-direction. 
\textbf{e}-\textbf{g}, Real and imaginary parts of the band structures for (\textbf{e}) the trivial phase with $\nu = 0.6$, (\textbf{f}) the gapless phase with $\nu = 1.5$, and (\textbf{g}) the nontrivial phase with $\nu = 2$. The band structures are calculated from the circuit lattices with PBCs in both the $x$- and $y$-directions. In panels (\textbf{b})-(\textbf{g}), the strength of non-Hermicity is set to $\delta \nu = 0.5$.
 }
\label{fig2}
\end{figure*}

Here, we report the realization of the non-Hermitian Haldane model in an electric circuit lattice. We utilize the second method for implementing the non-Hermitian model by breaking the antisymmetry of the next-nearest-neighbor (NNN) hopping. Our observations reveal amplifying and dissipative topological edge states, which are challenging to achieve on other platforms. We break both inversion symmetry and time-reversal symmetry, demonstrating that non-Hermiticity can induce the closure of the line gap, corresponding to a semimetal phase. Additionally, we present the robustness of these edge states. Our experimental findings pave the way for the exploration of unusual non-Hermicity-induced topological phases within the Haldane model.
\newline

\noindent \textbf{Results}

\noindent \textbf{Circuit structure.} Figures \ref{fig1}a and b illustrate the fabricated printed circuit board (PCB) and the schematic of the circuit that realizes the non-Hermitian Haldane model, respectively. The PCB consists of $6 \times 6$ unit cells; however, for simplicity, we only display $4 \times 4$ unit cells in Fig. \ref{fig1}b. The grounded $LC$ resonators with nodes A (red) and B (blue) emulate the on-site energies, with parameters set to $L_{\text{g}} = 68~\mu \text{H}$, $C_{\text{A}} = 20~\text{nF}$, and $C_{\text{B}} = 0$. The capacitors, each with a value of $C = 10~\text{nF}$, that connect the nodes correspond to the nearest-neighbor (NN) hopping. The negative impedance converters (NICs) facilitate the nonreciprocal next-nearest-neighbor (NNN) hopping with $\phi = \pm \pi/2$ in the Haldane model \cite{PRL122-247702,PRB99-235110}. By adjusting the values of the resistors in the NICs, we achieve $R_{\text{A},\text{B}} = -R/\nu_{2,1}$ for the left side and $R_{\text{A},\text{B}} = R/\nu_{1,2}$ for the right side of the NICs, where we set $R = 1~\text{k}\Omega$ (see Supplementary Information, Section 3). It is important to note that the time-reversal symmetry of this circuit lattice is explicitly broken by the nonreciprocal NNN hopping, which sharply contrasts with the pseudo-$T$ breaking employed in previous studies \cite{SCPMA64-257011,PRX5-021031}. For ease of fabrication and measurement, we deform the honeycomb lattice into a brick wall structure (see Supplementary Information, Section 1). Since this deformation is equivalent to a gauge transformation \cite{PRL122-247702,PRB99-161114}, the Hamiltonian of the deformed circuit lattice resembles that of the Haldane model. Furthermore, we implement periodic boundary conditions in the longitudinal $y$-direction by connecting the first node to the last node, while applying open boundary conditions in the horizontal $x$-direction by grounding.

\begin{figure*}[tbp]
\includegraphics[width=17.2cm]{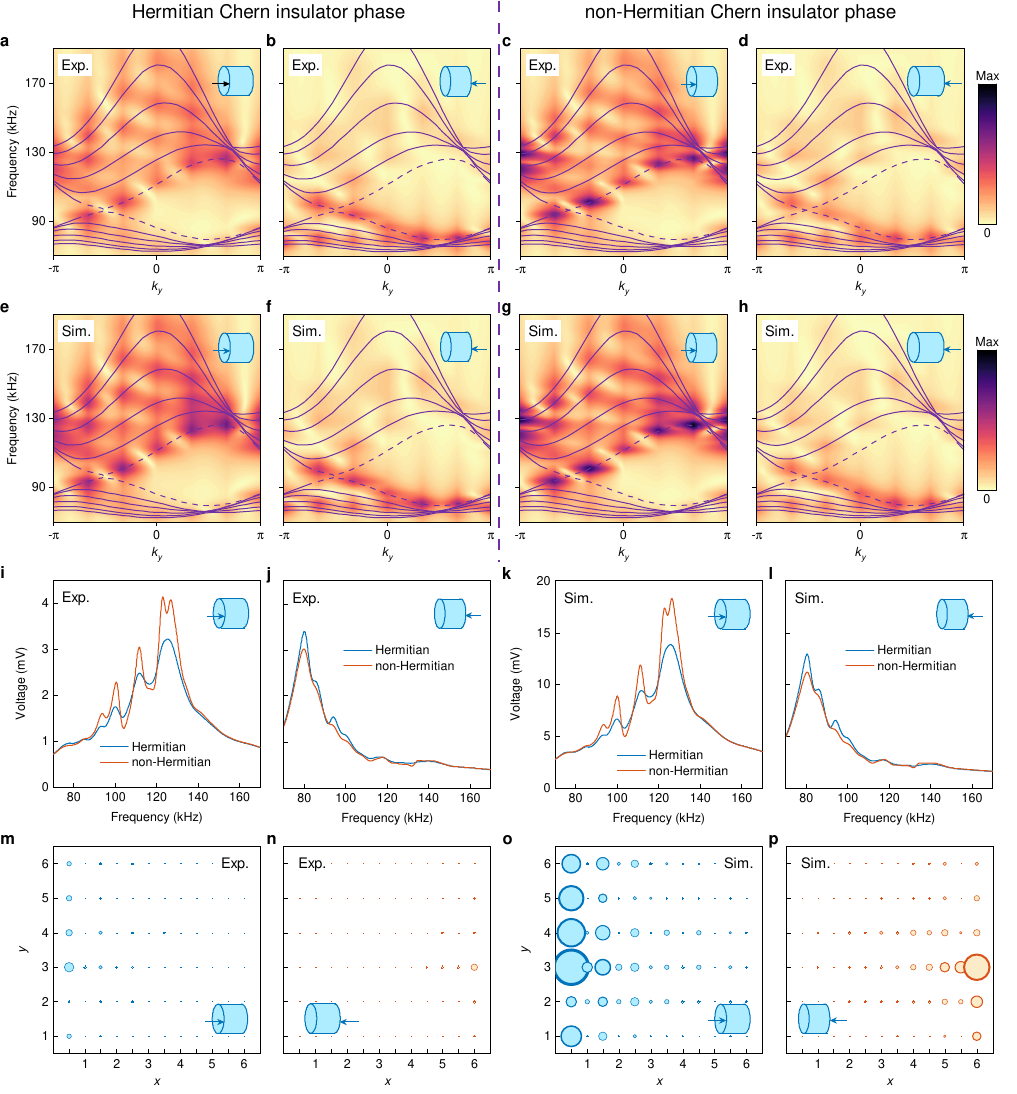}
\caption{\textbf{Band structures and voltage spectra for the Chern insulator phase.} 
\textbf{a}-\textbf{b}, Experimental band structures for the Hermitian Chern insulator phase with $\nu = 1.47$ and $\delta \nu = 0$. 
\textbf{c}-\textbf{d}, Experimental band structures for the non-Hermitian Chern insulator phase with $\nu = 1.47$ and $\delta \nu = 0.1$. 
\textbf{e}-\textbf{h}, Band structures obtained from simulation results. In panels (\textbf{a})-(\textbf{h}), the purple curves represent the theoretical band structures calculated with $N_{x} = 6$ and $N_{y} = 30$, where solid and dashed curves correspond to bulk and edge states, respectively. 
\textbf{i}-\textbf{l}, Voltage spectra for the Hermitian and non-Hermitian Chern insulator phases at the excitation nodes. 
\textbf{m}-\textbf{p}, Voltage distributions at (\textbf{m} and \textbf{o}) $f = 111.4~\text{kHz}$ and (\textbf{n} and \textbf{p}) $f = 86.3~\text{kHz}$, respectively. In panels (\textbf{a})-(\textbf{p}), the arrows in the insets indicate the excitation positions on the circuit sample.}
\label{fig3}
\end{figure*}

\noindent \textbf{Phase diagram.} We first investigate the phase diagram of the circuit lattice under periodic boundary conditions (PBCs) in both the $x$ and $y$ directions. We define $\nu = \left( \nu_{1} + \nu_{2} \right)/2$ and $\delta \nu = \left( \nu_{1} - \nu_{2} \right)/2$, which characterize the strength of $T$ breaking and the asymmetry of the NNN hopping (i.e., the strength of non-Hermiticity). When $\delta \nu = 0$, the circuit Hamiltonian is Hermitian. Due to the breaking of $P$ with $C_{\text{A}} \neq C_{\text{B}}$, a phase transition boundary exists between the trivial insulator and the Chern insulator, as illustrated in Fig. \ref{fig2}a. This boundary corresponds to a semi-metallic phase characterized by $\nu = \omega \left( C_{\text{A}} - C_{\text{B}} \right) R / 6\sqrt{3}$ (see Supplementary Information, Section 1). At this boundary, the frequency band structure exhibits a single Dirac point \cite{ncommun11-1873}. 
When $\delta \nu \neq 0$, the circuit Hamiltonian becomes non-Hermitian, and a regime characterized by the gapless line gap, defined by $\text{Re} \left( f_{0} \right) = 0$, appears in the phase diagram, where $f_{0} = \frac{1}{2\pi} \sqrt{L_{\text{g}} \left( C_{\text{g}} + 3C \right)}$ and $C_{\text{g}} = \left( C_{\text{A}} + C_{\text{B}} \right)/2$. Increasing $\delta \nu$ expands the gapless regime. 
In Figs. \ref{fig2}b-d, we present the complex frequency spectra for three distinct phases in the phase diagram with $\delta \nu = 0.5$. For simplicity, we assume $R_{\text{g}} = \infty$. In Figs. \ref{fig2}b and d, with $\nu = 0.6$ and $\nu = 2$, respectively, the frequency spectra both exhibit line gaps. However, in Fig. \ref{fig2}c, with $\nu = 1.5$, the line gap is closed, indicating the gapless regime shown in Fig. \ref{fig2}a. To differentiate the phases in Figs. \ref{fig2}b and d, we calculate their Chern numbers using biorthogonal eigenstates (see Supplementary Information, Section 2). We find that Fig. \ref{fig2}b corresponds to a trivial insulator with $C = 0$, while Fig. \ref{fig2}d corresponds to a nontrivial Chern insulator with $C = -1$. The regions between these two gapped phases represent semimetallic phases with gapless line gaps. 
To further validate the phase diagram in Fig. \ref{fig2}a, we also present the band structures for the three different phases in Figs. \ref{fig2}e-g. The real parts of the band structures in Figs. \ref{fig2}e and g exhibit bandgaps, while Fig. \ref{fig2}f is gapless. All imaginary parts are gapless, demonstrating that the state in Fig. \ref{fig2}f contains exceptional points. The six exceptional points are separated by the Dirac point at $\mathbf{K}=\left( \frac{2}{3} \pi, -\frac{2}{3} \pi \right)$ for a Hermitian Dirac semimetal, forming a loop in the $\left( k_{x}, k_{y} \right)$ plane (see Supplementary Information, Section 3). These results confirm the existence of line gaps characterized by $\text{Re} \left( f_{0} \right) = 0$ for both the trivial and Chern insulator phases, as well as gapless line gaps for the semimetal phases.

We then consider a stripe geometry with a B-A terminated open boundary in the $x$-direction and a periodic boundary condition in the $y$-direction. This circuit sample is equivalent to a cylinder with the axial direction aligned along the $x$-direction. From the complex frequency spectra shown in Figs. \ref{fig2}b-d, we observe that the frequencies for the bulk states form two distinct clusters (indicated in blue), and the frequency spectra align with the corresponding periodic-boundary spectrum (shown in red) on the complex plane.
Although there are line gaps between the two clusters for both the trivial insulator phase in Fig. \ref{fig2}b and the Chern insulator phase in Fig. \ref{fig2}d, edge states cross the line gap and connect the two clusters in the Chern insulator phase. Specifically, one edge state is amplifying while the other is dissipative, due to their opposite signs of the imaginary parts of the frequencies. In Fig. \ref{fig2}c, edge states still exist and connect the two clusters, even though this corresponds to a semimetal phase characterized by a gapless line gap.
It is important to note that there is no skin effect present throughout the entire phase diagram. This is because the net nonreciprocal pumping, which can lead to the skin effect, is canceled in this particular stripe geometry \cite{PRL123-016805}. As a result, both the bulk and edge states do not exhibit the skin effect. In the following, we will investigate the two previously uncovered phases: a non-Hermitian Chern insulator phase and a non-Hermitian semimetal phase.

\noindent \textbf{Non-Hermitian Chern insulator phase.} To investigate the non-Hermitian Chern insulator phase, we set $\nu = 1.47$ and $\delta \nu = 0.1$ (see Supplementary Information, Section 4). Since directly observing an amplifying edge state experimentally is challenging, we introduce grounded resistors with $R_{\text{g}} = 1~\text{k}\Omega$ at both nodes A and B. This modification transforms the amplifying and dissipative edge states into two dissipative edge states with unequal losses. We experimentally excite the circuit sample from both the left and right sides and measure the band structures and voltage spectra for the non-Hermitian Chern insulator phase (see Supplementary Information, Section 3). Under these excitations, the edge states propagating along the two edges are excited independently.
For comparison, Figs. \ref{fig3}a-b display the experimental band structures for a Hermitian Chern insulator phase with $\delta \nu = 0$. The purple curves represent the theoretical band structures calculated with $N_{x} = 6$ and $N_{y} = 30$, where solid and dashed curves correspond to bulk and edge states, respectively. The excitation positions on the circuit sample are indicated by arrows in the insets.

In Figs. \ref{fig3}c-d, we present the band structures for the non-Hermitian Chern insulator phase, where the edge states are located in the gap between the bulk states. The edge state in Fig. \ref{fig3}c is more pronounced, indicating that it has lower loss. Considering the lossy background induced by the grounded resistors $R_{\text{g}}$, the edge state in Fig. \ref{fig3}c acts as an amplifying edge state, while the edge state in Fig. \ref{fig3}d is a dissipative edge state. Furthermore, similar to the chiral edge states in a Hermitian Chern insulator phase, the edge states in the non-Hermitian case remain chiral, meaning they propagate in opposite directions.
Figs. \ref{fig3}e-h show the band structures obtained from simulation results, which align well with the experimental findings. The presence of amplifying and dissipative edge states is further confirmed by the voltage spectra measured at the excitation nodes. From Figs. \ref{fig3}i-l, the voltage spectrum for the amplifying edge state exhibits higher $Q$ factors, while the spectrum for the dissipative edge states is smoother than that of the Hermitian Chern insulator phase.
To illustrate the positions of the edge states, we also map the voltage distributions at $f = 111.4~\text{kHz}$ (Figs. \ref{fig3}m and o) and $f = 86.3~\text{kHz}$ (Figs. \ref{fig3}n and p), respectively. Thus, in the non-Hermitian Chern insulator phase, the edge state at the left edge is amplifying, while the edge state at the right edge is dissipative.
Additionally, a notable feature of edge states in a Hermitian Chern insulator is their robustness against lattice disorders and defects. In our circuit lattice, we introduce circuit defects by grounding several nodes along the edges (see Supplementary Information, Section 4). By comparing the voltage spectra for the non-Hermitian Chern insulator phases with and without defects, we find that both the amplifying and dissipative edge states exhibit remarkable robustness.

\begin{figure*}[tbp]
\includegraphics[width=17.2cm]{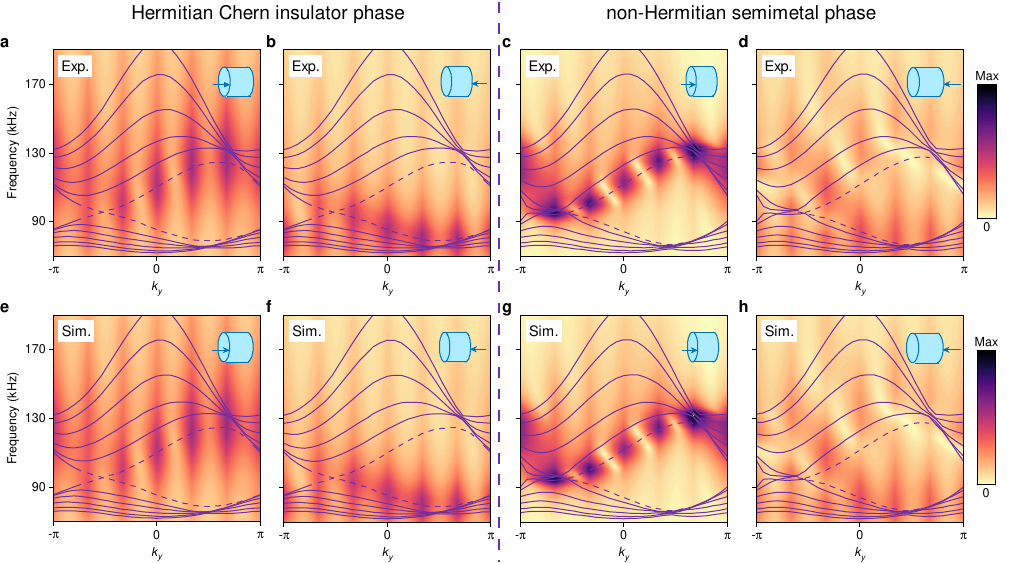}
\caption{\textbf{Band structures and voltage spectra for the semimetal phase.} 
\textbf{a}-\textbf{b}, Experimental band structures for the Hermitian Chern insulator phase with 
$\nu = 1.47$ and $\delta \nu = 0$. \textbf{c}-\textbf{d}, Experimental band structures for the non-Hermitian 
semimetal phase with $\nu = 1.47$ and $\delta \nu = 0.7$. \textbf{e}-\textbf{h}, Band structures retrieved from 
the simulation results. In (\textbf{a})-(\textbf{h}), the purple curves show the theoretical band structures with 
$N_{x} = 6$ and $N_{y} = 30$, where the solid and dashed curves correspond to the bulk and edge states, 
respectively. The arrows in the insets illustrate the excitation positions on the circuit sample.}
\label{fig4}
\end{figure*}

\noindent \textbf{Non-Hermitian semimetal phase.} The asymmetry of the next-nearest-neighbor (NNN) hopping is equivalent to the strength of non-Hermicity. For a Hermitian circuit Hamiltonian with $\delta \nu = 0$, the semimetal phase with a gapless band structure occurs only for a specific value of $\nu$. In contrast, in the non-Hermitian case, as shown in the phase diagram in Fig. \ref{fig2}a, increasing $\delta \nu$ expands the gapless regime. To demonstrate the non-Hermitian semimetal phase induced by non-Hermicity, we fabricate another circuit sample with $\nu = 1.47$ and $\delta \nu = 0.7$ (see Supplementary Information, Section 4), which corresponds to the semimetal phase regime.
To mitigate the increased amplification of the amplifying edge state, we utilize grounded resistors with $R_{\text{g}} = 160~\Omega$. Figs. \ref{fig4}a-d and Figs. \ref{fig4}e-h present the experimental and simulated band structures, respectively. For comparison, the band structures of a Hermitian Chern insulator phase with $\delta \nu = 0$ are shown in Figs. \ref{fig4}a-b and Figs. \ref{fig4}e-f.
From Figs. \ref{fig4}c-d and Figs. \ref{fig4}g-h, we observe that the frequency gap separating the bulk states is closed, indicating the presence of the non-Hermitian semimetal phase. The chiral edge states continue to exist in this semimetal phase. Due to the increased non-Hermicity, the edge state at the left edge is significantly amplified, as illustrated in Figs. \ref{fig4}c and g. In contrast, the edge state at the right edge is barely distinguishable due to substantial loss, as shown in Figs. \ref{fig3}d and h. Furthermore, both the left and right edge states remain robust against lattice defects (see Supplementary Information, Section 5). \newline

\noindent \textbf{Conclusion.} We have experimentally realized the genuine non-Hermitian Haldane model in an electric circuit lattice. By tuning the NNN
 hopping to break Hermiticity, we demonstrate two previously uncovered phases in the Haldane model: a non-Hermitian Chern insulator phase and a non-Hermitian semimetal phase. Both phases exhibit simultaneously amplifying and dissipative topological edge states, albeit at different boundaries. Our results not only represent the first experimental realization of the non-Hermitian Haldane model but also pave the way for exploring unusual non-Hermicity-induced topological phases. Looking ahead, it would be intriguing to investigate other non-Hermicity-induced phenomena, such as exceptional points and the non-Hermitian skin effect, in the non-Hermitian Haldane model with appropriately designed couplings.  \newline

%\noindent \textbf{Author contributions}
%
%\noindent R.L. conceived the idea. R.L. and W.W. performed the theoretical calculations
%and simulations. R.L., W.W., X.K., B.L., Y.J., H.T. designed and conducted the experiments. 
%R.L., H.T., and Y.L. supervised the project. \newline

\noindent \textbf{Data availability statement} 

\noindent The data that support the findings of this study are available from the 
corresponding author upon reasonable request. \newline

\noindent \textbf{Acknowledgements}

\noindent The authors thank fruitful discussions with Baile Zhang and Yihao Yang.
R.L., W.W., and X.K. was sponsored by the National Key Research and Development
Program of China (Grant No. 2022YFA1404902), National Natural Science
Foundation of China (Grant No. 12104353), and the Open Foundation of the 
State Key Laboratory 
of Modern Optical Instrumentation. B.L. was sponsored by the
National Natural Science Foundation of China under Grant No. 61901133; 
Fundamental Research Funds for the Central Universities (3072024XX2504); 
Forward Design Technology Special Fund Project of Harbin Engineering University
(KYWZ220242504, KYW220240807, KYWZ220240807).
P.L. was sponsored by the
National Natural Science Foundation of China (11805141) and
Basic Research Program of Shanxi Provence (202203021222250).
Y.L. was sponsored by the National 
Natural Science Foundation of China (NSFC) under Grant No. 62271366 and the 
111 Project. \newline

\noindent \textbf{Competing interest} 

\noindent The authors declare no competing financial interest.

\end{document}